\documentclass[aps,float,prl,psfig]{revtex4}
\input epsf
\usepackage{graphicx}
\newcommand{\beq}{\begin{equation}}
\newcommand{\beqa}{\begin{eqnarray}}
\newcommand{\eeq}{\end{equation}}
\newcommand{\eeqa}{\end{eqnarray}}

\newcommand{\siml}{\lesssim}

\newcommand{\vect}[1]{\mbox{\boldmath${#1}$}}
\newcommand{\lmk}{\left(}
\newcommand{\rmk}{\right)}

\newcommand{\lkk}{\left[}
\newcommand{\rkk}{\right]}
\newcommand{\lla}{\left\langle}

\newcommand{\rra}{\right\rangle}

\newcommand{\vex}{{\vect x}}

\newcommand{\vel}{\vect l}

\newcommand{\ven}{\vect n}
\newcommand{\vem}{\vect m}

\newcommand{\ve}{{\vect e}}

\begin{document}
\title{
Prospects for direct detection of  circular polarization of 
    gravitational-wave background
} 
\author{Naoki Seto
}
\affiliation{Department of Physics and Astronomy, 4186 Frederick Reines
Hall, University of California, Irvine, CA 92697
}

\begin{abstract}

We  discussed prospects for directly detecting  circular 
polarization signal of gravitational wave background. We  found it is
generally difficult to probe the  monopole mode of the signal due to
broad directivity of gravitational wave detectors. 
But the dipole  $(l=1)$ and octupole $(l=3)$ modes of the signal can be
measured in a  simple manner by combining outputs of two unaligned
detectors, and we can dig them deeply under confusion  and detector
noises.   Around $f\sim 0.1$mHz LISA will provide ideal data
 streams to detect these anisotropic components whose magnitudes are as
 small as  $\sim1$ percent of  the detector  noise level  in 
 terms  of the non-dimensional  energy 
density $\Omega_{GW}(f)$. 
\end{abstract}
\pacs{PACS number(s): 95.55.Ym 98.80.Es, 95.85.Sz }
\maketitle

\section{introduction}
As gravitational interaction is very weak,  significant efforts have
been made 
to detect gravitational waves. But, on the other hand, we will be able
to get rich 
information of the universe by observing gravitational waves that
directly propagate 
to us with almost no absorption. Various astrophysical and cosmological
models predict existence of stochastic gravitational wave background,
and it is an interesting target for  gravitational wave astronomy
\cite{Allen:1996vm}. For  
its observational prospects, we need to understand how we characterize the background 
 and what aspects we can
uncover  with current  and future observational facilities. 

One of such aspects is circular polarization that describes whether the
background has asymmetry with respect to magnitudes of right- and
left-handed waves. Circular polarization of gravitational wave
background might be generated by helical 
turbulent motions (see {\it e.g.} \cite{Kahniashvili:2005qi} and
references therein). Inflation
scenario predicts gravitational wave background 
from quantum fluctuations during acceleration phase in the early
universe, but asymmetry of left and right-handed waves can be produced
with the gravitational Chern-Simon term that might be derived from
string theory and might be related to
creation of baryon number (see {\it e.g.} \cite{Lyth:2005jf} and
references therein). Primordial gravitational wave background, including
information of its 
circular polarization \cite{Lue:1998mq},  can be indirectly
studied with CMB measurement \cite{Seljak:1996gy} at very low frequency
regime $f\sim 
10^{-17}$Hz that is largely different from the regime directly
accessible with 
gravitational wave detectors studied in this Letter. Gravitational wave
background from 
Galactic binaries can be polarized, if orientation of their angular
momentum have coherent distribution, such as, correlation with Galactic
structure.  While  observational samples of local Galactic binaries
 do not 
favor such  correlation \cite{inc}, this will also be an interesting
target that can be directly studied with  LISA.

It is well known that observation of gravitational wave is
intrinsically sensitive to its polarization state \cite{Thorne_K:1987}. 
 This is because we measure spatial expansion and
contraction due to the wave, and 
polarization determines the direction of the oscillation perpendicular
to its propagation direction. Another 
important nature of the observation is that we have to simultaneously
deal with waves basically coming from all the directions, in contrast
to observations with typical electromagnetic wave telescopes that have
sharp directivity. Therefore, polarization information and directional
information couple strongly in observational analysis of gravitational
wave background. In general, orientation of a gravitational wave
detector  
changes with time, and induced  modulation of data stream is useful to
probe the polarization and directional information. For ground  based detectors, such as, LIGO, this is 
due to the daily rotation of the Earth \cite{Allen:1996gp}. For a space mission like LISA,
this 
change is determined by its orbital choice \cite{lisa,Cornish:2001qi}. In addition,  we
can also 
expect that 
several independent data streams of gravitational waves will be taken at
the same time 
\cite{Cutler:1997ta,Prince:2002hp}. In this 
Letter, in view of these observational characters,  we study how well we
can 
extract information  of circular 
polarization of the background 
in a model independent manner about its origin.   
\section{formulation}

The standard plane wave expansion of metric perturbation by
gravitational waves is given as 
\beq
h_{ab}(t,\vex)=\sum_{P=+,\times} \int^{\infty}_{-\infty} df \int_{S^2} d\ven
h_P(f,\ven) e^{2\pi i f (t-\ven \cdot \vex) } \ve^P_{ab}(\ven),
\eeq
where $S^2$ is the unit sphere for the angular integral, the unite vector
$\ven=(\sin\theta\cos\phi, \sin\theta\sin\phi,\cos\theta)$ is  
 propagation direction, and  $\ve^+_{ab}$ and $\ve^\times_{ab}$ are the basis for the transverse-traceless tensor.  We fix them as
$\ve^+_{ab}={\hat \ve}_\theta \otimes {\hat \ve}_\theta- {\hat \ve}_\phi
\otimes  {\hat \ve}_\phi$ and $\ve^\times_{ab}={\hat \ve}_\theta \otimes
{\hat 
\ve}_\phi+{\hat  
\ve}_\phi \otimes {\hat \ve}_\theta$ where ${\hat \ve}_\theta$ and ${\hat
\ve}_\phi$ are two unit vectors with a fixed spherical coordinate
system.     
As the metric perturbation $h_{ab}(t,\vex)$ is real, we have a relation
for complex conjugate;
$h_P(-f,\ven)=h_P(f,\ven)^*$.
When we replace the direction $\ven\to -\ven$, the matrices have
correspondences $ \ve^+_{ab} \to \ve^+_{ab}$ (even parity) and
$\ve^\times_{ab}\to 
-\ve^\times_{ab}$ (odd parity).   

To begin with, we  study   gravitational wave modes at a 
frequency $f$.  Here,
we omit  explicit frequency 
dependence for notational simplicity, unless we need to keep it. The
covariance matrix 
$\left( 
           \begin{array}{@{\,}cc@{\,}}
           \lla h_+(\ven) h_+^* (\ven') \rra & \lla h_+ (\ven)
	    h_\times^* (\ven') \rra  \\
            \lla h_+^*(\ven)  h_\times(\ven')\rra & \lla h_\times(\ven)
	      h_\times^*(\ven')\rra \\  
           \end{array} \right)
$ 
for two polarization modes $h_+(\ven)$ and $h_\times(\ven)$
 is decomposed  as 
\beq
\frac{\delta_{drc}(\ven-\ven')}{4\pi}\left( 
           \begin{array}{@{\,}cc@{\,}}
           I+Q & U-iV  \\
            U+iV & I-Q  \\ 
           \end{array} \right),
\eeq
where the symbol $\lla \dots\rra$ represents to take an ensemble average
for superposition of stationary incoherent waves, and
$\delta_{drc}(\cdot)$ is  the   delta 
function on the unit sphere $S^2$. 
We have defined the following Stokes parameters \cite{radipro} analog to
electromagnetic 
waves 
as   
$
I(\ven)= \lla |h_+|^2+|h_\times|^2\rra /2, ~~Q(\ven)=\lla |h_+|^2-|h_\times|^2\rra
/2, ~~U(\ven)= \lla h_+ h_\times^*+ h_+^* h_\times \rra /2, $ and
$V(\ven)=i \lla h_+ 
h_\times^*- h_+^* h_\times \rra /2.
$
The parameter $I(\ven)$ represents the total intensity of the wave, while $Q(\ven)$
and $U(\ven)$ is related to linear polarization.  The parameter $V(\ven)$ is related
to circular 
polarization, and its sign shows whether right- or left-handed
waves dominate.  When we rotate the basis vectors ${\hat \ve}_\theta
$ and $ 
{\hat \ve}_\phi$ around the axis $\ven$ by angle $\psi$, the parameters
$I$ 
and $V$ are invariant (spin 0). But the parameters $Q$
and $U$ are transformed as 
$
(Q\pm iU)'(\ven)=e^{\mp 4i \psi}(Q\pm iU)(\ven),
$
and the combinations $(Q\pm iU)(\ven)$ have spin $\pm 4$
\cite{radipro}. 

Next we discuss  response of a two-arm interferometer $J$ with
$90^\circ$ vertex angle and equal arm-length $L$. We put $\vel_1$ and
$\vel_2$ as unit vectors for the directions of the arms.  The beam pattern
functions $F_J^P=(\vel_1\cdot \ve^P\cdot \vel_1-\vel_2\cdot \ve^P\cdot \vel_2)/2$ $(P=+,\times)$
represent relative sensitivities to two linearly polarized gravitational
waves 
 ($P=+,\times$) with various directions $\ven$
\cite{Thorne_K:1987}. Note that the function $F_J^+$ has even 
parity 
and $F_J^\times$ has odd parity with respect to the 
direction $\ven$ (see {\it e.g.} \cite{Kudoh:2004he}). 

In  this Letter we mainly deal with  low frequency regime with
$f/f_*\ll 1$. Here, using the arm-length $L$, we have  defined a characteristic frequency
$f_*=1/(2\pi 
L)$ that corresponds to 10mHz for LISA and 1Hz for Big Bang Observer
(BBO) \cite{bbo,Seto:2005qy}. 
LISA is formed by three spacecrafts that nearly keep a regular  triangle
configuration \cite{lisa}. From its six one-way data streams, we can make
Time-Delay-Interferometer (TDI) variables that cancel laser frequency
noises. We can select three  TDI variables $A$, $E$ and $T$
whose detector noises are not correlated and can be regarded as
independent \cite{Prince:2002hp}. At the low frequency regime the
responses of the $A$ and $E$ modes can be effectively regarded as those of
two-arm interferometers whose configuration are shown in figure 1
\cite{Cutler:1997ta}. In 
this figure we
put the whole system on $XY$-plane ($\theta=\pi/2$).  The beam pattern
functions of the  $A$
mode are explicitly given as  
\beq
F_A^+(\theta,\phi)=\frac12  (1+\cos^2\theta) \cos2\phi,~~~
F_A^\times(\theta,\phi)=-\cos\theta \sin2\phi,\label{f2}
\eeq
while those for the $E$ mode are given by replacing $\phi\to\phi-\pi/4$
in the 
above expressions. 
The beam pattern functions for the $T$ mode are quite different from the
$A$ 
and 
$E$ modes, and  given as $F_T^P=\sum_{i=1}^3 (\vem_i\cdot \ve^P\cdot
\vem_i)(\vem_i\cdot \ven)$  $(P=+,\times)$ where directions of three unit
vectors 
$\vem_i$ are shown in figure 1  \cite{Kudoh:2004he,Tinto:2001ii}. At low
frequency regime 
sensitivity of the $T$ mode to gravitational waves is $\sim (f/f_*)^{-1}$ times worse than those
for the $A$ and $E$ modes \cite{Prince:2002hp,Tinto:2001ii}. Therefore we put this mode aside for a while. 

We can  express  responses of the $A$ and $E$ modes to gravitational
waves from a single direction $\ven$ as $(J=A,E)$
\beq
r_J(\ven) = \lmk F_J^+  h_+ (\ven)+ F_J^\times h_\times
(\ven) \rmk \lkk 
1+i(f/f_*)D_J(\ven)+O(f^2/f_*^2)\rkk, \label{ha}
\eeq
where the second factor $\lkk\dots  \rkk$ is  corrections caused by
the 
finiteness  
of the arm-length, and two  real functions $D_A(\theta,\phi)$ and
$D_E(\theta,\phi)$ depend on the propagation directions $\ven$ of waves
(see {\it e.g.} \cite{Kudoh:2004he}).
In eq.(\ref{ha})  we neglected 
overall factors that depend only on  frequency $f$ and are irrelevant for
our 
study. These formal expressions for the perturbative expansion with
respect to the ratio $(f/f_*)$ can be generally used with relevant beam
pattern functions, including the
case for the  $T$-mode or Fabri-Perot detectors as LIGO.


The information $V$  cannot be produced from the response $r_A$ or $r_E$
alone. 
To get it
 we  need independent linear combination of $h_+$ and
$h_\times$. 
For example  the term $ \lla r_A(\theta,\phi) r_A(\theta,\phi)^* \rra$
can 
be 
written only with $I,Q$ and $U$.  We now take
the low frequency limit, 
and keep only the leading order terms of the expansions.  The cross term
$\lla r_A(\ven)  
r_E(\ven')^*\rra$ is given as 
\begin{widetext}
\beqa
\lla  r_A(\ven) r_E(\ven')^*\rra
&=&\frac{\delta_{drc}(\ven-\ven')}{4\pi}{\Bigg\{}\frac12 \lkk\lmk 
\frac{1+\cos^2\theta}2 
\rmk^2   -\cos^2\theta
  \rkk \sin4\phi 
 I(\theta,\phi)
   + \frac12\lkk\lmk \frac{1+\cos^2\theta}2
\rmk^2   +\cos^2\theta
  \rkk  \sin4\phi
 Q(\theta,\phi)\nonumber\\
& & + \lkk\lmk \frac{1+\cos^2\theta}2
\rmk   \cos\theta
  \rkk \cos4\phi   
 U(\theta,\phi)
-i \lkk\lmk \frac{1+\cos^2\theta}2
\rmk   \cos\theta
  \rkk   
 V(\theta,\phi) {\Bigg\}}.\label{corr}
\eeqa
\end{widetext}
With the following combination 
\beq
 {\rm Im}\lkk \lla r_A(\ven) r_E(\ven')^* \rra \rkk= - \frac{\delta_{drc}(\ven-\ven')}{4\pi} \lkk\lmk
\frac{1+\cos^2\theta}2 
\rmk   \cos\theta
  \rkk   V(\theta,\phi)\label{defs}
\eeq 
we can extract  the circular parameter $V$ alone. 
 As shown in figure 1, the effective detector $E$ is obtained by
rotating the detector $A$ around $Z$-axis by $\pi/4$
\cite{Cutler:1997ta}. If this angle is 
$\delta$, 
there appears a factor $\sin 2\delta$ in the final expression in
eq.(\ref{defs}). Therefore, in some sense,   LISA will provide an optimal set $(A,E)$ to
study the parameter $V$.
In contrast,  sensitivity of correlation analysis to the monopole
intensity $I_{00}$  is 
proportional to $\cos 2\delta$, and LISA cannot probe it with the
method, as is well known.

We now discuss responses of  detectors to gravitational waves
from all  directions 
$\ven$.  
Firstly,  what we can get observationally form the $A$ and $E$ modes
are the following integrals 
$R_A=\int r_A(\ven)d\ven$ and $R_A=\int r_A(\ven)d\ven$. 
Considering the spin of quantities $I$, $Q\pm iU$ and $V$ we can
expand them in terms of the spin weighted spherical harmonics ${}_s Y_{lm}$
as  
$
I(\theta,\phi)=\sum I_{lm}Y_{lm}(\theta,\phi),~~~(Q\pm
iU)(\theta,\phi)=\sum K_{\pm ,lm}\cdot{}_{\pm 4}Y_{lm}(\theta,\phi),~~~V(\theta,\phi)=\sum V_{lm}Y_{lm}(\theta,\phi), 
$
where the coefficients $K_{\pm ,lm}$ are defined  for $l\ge 4$.
With these harmonic expansions and eq.(\ref{defs}), the combination
 ${\cal C}={\rm Im} \lla R_A R_E^*\rra$ is evaluated as
$
{\cal C}=-\frac85\sqrt{\frac{\pi}3}V_{10}-\frac25\sqrt{\frac{\pi}7}V_{30}.
$
When we rotate two interferometers $A$ and $E$ with Euler angles
$(\alpha,\beta,\gamma)$, the signal ${\cal C}$ becomes
\beq
{\cal C}(\alpha,\beta,\gamma)=-\frac{16\pi}{15}V_{1m}Y_{1m}(\alpha,\beta)-\frac{4\pi}{35}V_{3m}Y_{3m}(\alpha,\beta). \label{defsa}
\eeq 
This result is obtained by relating the coefficients $V_{lm}$  in
original coordinate with those in  rotated coordinate
\cite{Allen:1996gp}. 
LISA moves around the Sun with changing the orientation of its detector
plane that is 
inclined to  the 
ecliptic plane by 60 degree \cite{lisa}. To describe these motions in
the ecliptic  
coordinate, the Euler
angles  are given as
$\alpha=2\pi (t/1{\rm yr})+\alpha_0$, $\beta=\pi/3$ and $\gamma=-2\pi
(t/1{\rm yr})+\gamma_0$ with time $t$ and constants $\alpha_0$ and
$\gamma_0$ \cite{Cutler:1997ta}. This parameterization is also valid for
BBO \cite{bbo}. The 
parameter $\gamma$ determines  the
so-called cartwheel motion, but the combination ${\cal
C}(\alpha,\beta,\gamma)$ 
does not depend  on it.  From eq.(\ref{defsa}) we can understand that
  the observed signal  ${\cal C}(\alpha,\beta,\gamma)$ with LISA is
decomposed 
into  modulation patterns with frequencies $f=0,
\pm1/3,\pm1/2$ and $\pm1{\rm yr^{-1}}$  from $m=0,\pm1,\pm2$ and  $\pm 3$ modes respectively.

\begin{figure}
  \begin{center}
\epsfxsize=8.cm
\begin{minipage}{\epsfxsize} \epsffile{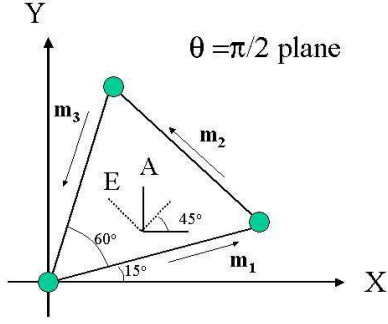} \end{minipage}
 \end{center}
  \caption{ The effective two-arm interferometers corresponding to the
 TDI modes $A$ and $E$. Three spacecrafts of LISA are shown with
 circles that are put on the $XY$-plane ($\theta =\pi/2$). The beam
 pattern functions for $T$ mode are given with three unit
 vectors $\vem_i$ as  $F_T^P=\sum_{i=1}^3 (\vem_i\cdot \ve^P\cdot
\vem_i)(\vem_i\cdot \ven)$ $(P=+,\times)$.}
\end{figure}

So far we have used the low frequency  approximation with $f/f_*\ll 1$.
When we increase the frequency $f$,  the correction terms
$i(f/f_*)D_A(\theta,\phi)$ and  $i(f/f_*)D_E(\theta,\phi)$ 
in eq.(\ref{ha})  change the  phases  of
$r_A(\theta,\phi)$ and 
$r_E(\theta,\phi)$ as a function of propagation directions $\ven$. With
these 
correction 
terms,  the combination $R_A R_E^*-R_A^* R_E$  has contributions of 
$I$, $Q$ and $U$ modes, and we cannot extract the circular polarization
$V$ in a clean manner. Note also that the first order term $O(f/f_*)$ for  the
combination $R_A R_E^*+R_A^* R_E$ depends on the parameter $V$ due to  the corrections. We can
restate the situation as follows; Roughly  
speaking, 
 the circular polarization is measured by correlating two
data with phase difference $\pi/2$ \cite{radipro}. This is related to
the fact  
$V(\ven)\propto {\rm Re}\lla  r_A(\theta,\phi)r_E(\theta,\phi )^* e^{i \pi/2}\rra$.
But the finiteness of the arm-length modulates the phase as a function of
direction, and we cannot keep the phase difference $\pi/2$ simultaneously
for all the 
directions. When we
consider the spatial separation ${\vect d}$ of two interferometers, same
kind of  arguments hold by perturbatively expanding  the phase
difference 
$\exp[2\pi i f {\vect d}\cdot \ven]$ with a expansion parameter $(f|{\vect d}|)$ in addition to that with $(f/f_*)$ for the effects of arm-length.  Therefore, in our 
analysis, the requirement for  the  low frequency 
regime 
is not for simplicity of calculation, but is a crucial condition to
extract the 
circular polarization alone with gravitational wave detectors that have
broad directional response and are remarkably
different from electromagnetic wave telescopes.

As shown in eq.(\ref{defs}), we cannot measure the monopole moment
$V_{00}$ 
of circular polarization using the signal $\cal C$ made from the $A$ and
$E$ 
modes  due to  parity reason at low frequency limit. More specifically,
 products, such as 
$F_A^+ F_E^\times$, 
have  odd parity. This simple results hold for signal $\cal C$ 
with any two two-arm interferometers at low frequency limit. It does not matter whether their
vertex angles are not $\pi/2$, whether two arm-lengths of each detector
are not 
equal, or whether two interferometers are on unparallel  planes
({\it e.g.} for LIGO and VIRGO combination). 
The signals $\cal C$
with  ($A,T$) or ($E,T$)  modes   depend on the coefficients $V_{lm}$
with even 
$l$ at their leading order. But, in the case of LISA, the combinations
do not have monopole 
moment $V_{00}$. This is because of the apparent symmetry 
of these data  streams \cite{Prince:2002hp,Kudoh:2004he}.   Furthermore we
cannot get the moment $V_{00}$ 
even using their higher order terms with 
$O((f/f_*)^n)$ $(n\ge 1)$, as long as LISA is symmetric at each vertex.
A future mission might use multiple LISA-type sets with data streams
$\{A,E,T\},  \{A',E',T'\},...$ \cite{bbo}. We confirmed  that even if detector
planes for $A$ and $T'$ modes are not parallel, the monopole $V_{00}$ can
not be captured by the signal $\cal C$ with their combination at their
leading order.

As we discussed so far, it is not straightforward to capture the monopole
$V_{00}$ in a simple manner. But, in principle, we can manage this. For
example, we add a detector  $E_2$ that is given by
moving the original detectors $E$ in figure 1 by distance $d$
toward $+
Z$-direction, and then take the signal $R_{A}R_{E_2}^*$.
At order $O((fd)^1 (f/f_*)^0)$, we can probe the monopole
$V_{00}$ \cite{prep}.
 This kind of
arrangement for  detector configuration might be important to study
anisotropy and polarization of  gravitational wave background in the
long run.  

\section{ observation with LISA}

Next we discuss how well we can analyze the information $V_{lm}$ of 
circular polarization of stochastic gravitational wave background with
LISA.  
The observed inclinations of local Galactic binaries are known to be
consistent with having random distribution with no correlation to global
Galactic  
structure \cite{inc}. This suggests that the confusion background by
galactic 
binaries is not polarized. In future, LISA itself will provide basic
parameters for thousands of Galactic binaries at $f\ge 3$mHz,  including
information 
of their
orientations \cite{lisa}.  This can be used to further  constrain the
allowed 
polarization degree of the Galactic confusion noise. 
While this information will help us to discriminate the origin of
 detected polarization signal with LISA, our analysis below does not
depend on this outlook.
Our target here is a combination $p\Omega_{GW,b}(f)$ where
$\Omega_{GW,b}(f)$ 
is the   
normalized energy density of gravitational wave background (see
\cite{Allen:1996vm} for its definition), regardless
of its origin. The parameter
$p=(\sqrt{\sum_m |V_{1m}|^2+\sum_m |V_{3m}|^2})/I_t$  is its circular
polarization power in $l=1$ and 3 multipoles ($I_t \equiv
\sqrt {\sum_{lm} 
|I_{lm}|^2}$ : total intensity). For simplicity, we do not
discuss technical aspects in relation to the time modulation of the
signal  ${\cal C}$
due to  the motion of LISA that was
explained earlier around eq.(\ref{defsa}). We can easily make an
appropriate extension  to deal with it \cite{Cornish:2001qi}.

As we will see below,  the signal $\cal C$ is a powerful probe to study
the target $p\Omega_{GW,b}(f)$ in a frequency regime
where the amplitude
$p\Omega_{GW,b}(f)$ itself may be dominated by the confusion background noise or  detector noise. 
We take a summation of the signal $\cal C$ for Fourier modes around a
frequency $f$ 
in a bandwidth $\Delta f\sim f$. As the frequency resolution is inverse
of the observational time $T_{obs}$ and the total number of Fourier
modes in 
the band is $(T_{obs} \Delta f)$, the expectation value for the
summation becomes ${\cal C} (T_{obs} \Delta f)\sim p I_t (T_{obs}\Delta f)$. 
Here we used the fact that the detector noises for the $A$ and $E$ modes are not correlated. 
In contrast the
fluctuation $N$ 
for this summation   is
given as 
$
N=\sum_f \lmk n_A(f)n_E(f)^*-n_A(f)^*n_E(f)    \rmk,
$
where the total noises $n_A(f)$ and $n_E(f)$ include both detector noise and
circularly unpolarized potion of the  confusion noise.
 The root-mean-square value  of $N$ is written as  
$
\lla |N|^2 \rra^{1/2}\sim S_n(f)(T_{obs }\Delta f)^{1/2}
$
with $S_n(f)$ being the total noise spectrum for the $A$ and $E$ modes.
Thus the signal to noise ratio for the measurement is $SNR\sim p I_T 
(T_{obs }\Delta f)^{1/2}/S_n(f)$. We can rewrite this expression with using 
the normalized energy density $\Omega_{GW}$ and obtain
$
SNR\sim \lmk \frac{p \Omega_{GW,b} }{\Omega_{GW,n}} \rmk  (T_{obs}\Delta
f)^{1/2}   ,
$
where the magnitude ${\Omega_{GW,n}}$ corresponds to the total noise
level around frequency $f$. 
When the detector noise is dominated by the background noise, we have
$\Omega_{GW,n}\sim \Omega_{GW,b}$. This will be the case for  LISA
around $0.3 {\rm mHz} \siml f \siml 2$mHz.
As a concrete example,  we put  $f\sim 0.1$mHz with the expected noise level 
${\Omega_{GW,n}}\sim 
10^{-10}$ \cite{lisa}, and take  bandwidth
$\Delta f \sim f$. Then we have
$
SNR\sim \lmk \frac{p \Omega_{GW,b} }{10^{-12}} \rmk  (T/3{\rm
yr})^{1/2}.   
$ 
The improvement factor  $(T_{obs}\Delta
f)^{-1/2}$ for the detectable level of the target $p \Omega_{GW,b}$ is caused by the same reason as standard
correlation technique
for detecting  stochastic background
\cite{Flanagan:1993ix}. Interestingly, this factor 
$(T_{obs}\Delta f)^{-1/2}\sim 0.01$ around $f\sim 0.1$mHz is almost same
as the 
maximum
level of relative  contamination $(f/f_*)\sim 0.01$ for LISA by other
modes ($I$, 
$Q$,
$U$) due to  finiteness of  arm-length. 
A Japanese future project DECIGO \cite{Seto:2001qf} is planed to use
Fabri-Perot type 
design with its characteristic sensitivity $f_*= 50$Hz \cite{Kawamura},
while its best 
sensitivity is around $\sim 0.3$Hz similar to BBO whose characteristic
frequency is  $f_*=1$Hz.
Therefore, DECIGO is expected to be less affected by other modes and has
potential to  
reach the level $p \Omega_{GW,b} \sim 10^{-16}$ with one year observation.

If  circular polarization parameter $V({\theta,\phi})$ of cosmological
background is highly isotropic in the 
CMB-rest frame, its dipole pattern will be induced by  our peculiar
motion \cite{Kudoh:2005as}. This might be an interesting observational
target in future. In this case 
we can predict 
the  time modulation pattern ${\cal
C}(\alpha,\beta,\gamma)$  for  rotating detectors. A consistency
check using this modulation will  
be powerful approach to discriminate its cosmological origin. Actually a
similar 
method can be 
used for   LISA to  analyze circular  polarization of Galactic
confusion noise.

The author would like to thank A. Taruya for comments, and
A. Cooray for various  supports.
This research was funded by McCue Fund at the Center for Cosmology,  
University of California, Irvine.




\begin{thebibliography}{DUM}


\bibitem{Allen:1996vm}
  M.~Maggiore,
  Phys.\ Rept.\  {\bf 331}, 283 (2000).


\bibitem{Kahniashvili:2005qi}
  T.~Kahniashvili, G.~Gogoberidze and B.~Ratra,
  Phys.\ Rev.\ Lett.\  {\bf 95}, 151301 (2005).


\bibitem{Lyth:2005jf}
S.~H.~S.~Alexander, M.~E.~Peskin and M.~M.~Sheikh-Jabbari,
  Phys.\ Rev.\ Lett.\  {\bf 96}, 081301 (2006).



\bibitem{Lue:1998mq}
  A.~Lue, L.~M.~Wang and M.~Kamionkowski,
  Phys.\ Rev.\ Lett.\  {\bf 83}, 1506 (1999).


\bibitem{Seljak:1996gy}
  U.~Seljak and M.~Zaldarriaga,
  Phys.\ Rev.\ Lett.\  {\bf 78}, 2054 (1997);
M.~Kamionkowski, A.~Kosowsky and A.~Stebbins,
  Phys.\ Rev.\ Lett.\  {\bf 78}, 2058 (1997).




\bibitem{inc}S.~S. Huang and C. Wade, Astrophys.J, {\bf 143},
 146 (1966).




\bibitem{Thorne_K:1987}
K.~S. Thorne,  in {\em Three hundred years of gravitation}, (Cambridge University Press, Cambridge, 1987),
  pp.\ 330--458.



\bibitem{Allen:1996gp}
  B.~Allen and A.~C.~Ottewill,
  Phys.\ Rev.\ D {\bf 56}, 545 (1997).




\bibitem{lisa}
P.~L.~Bender  { et al},
{\it LISA Pre-Phase A Report,} Second edition, July 1998. 


\bibitem{Cornish:2001qi}
  N.~J.~Cornish and S.~L.~Larson,
  Class.\ Quant.\ Grav.\  {\bf 18}, 3473 (2001);
C.~Ungarelli and A.~Vecchio,
  Phys.\ Rev.\ D {\bf 64}, 121501 (2001);
N.~Seto,
  Phys.\ Rev.\ D {\bf 69}, 123005 (2004);
A.~Taruya and H.~Kudoh,
  Phys.\ Rev.\ D {\bf 72}, 104015 (2005).

\bibitem{Cutler:1997ta}
  C.~Cutler,
  Phys.\ Rev.\ D {\bf 57}, 7089 (1998).



\bibitem{Prince:2002hp}
 T .~A.~Prince, et al,
  Phys.\ Rev.\ D {\bf 66}, 122002 (2002).



\bibitem{radipro}
G. B. Rybicki and A. P. Lightman, {\it Radiative Process in Astrophysics} (Wiley, New York, 1960). 

\bibitem{Kudoh:2004he}
  H.~Kudoh and A.~Taruya,
  Phys.\ Rev.\ D {\bf 71}, 024025 (2005).




\bibitem{bbo}
E. S. Phinney et al. The Big Bang Observer, NASA Mission Concept Study
 (2003).


\bibitem{Seto:2005qy}
  N.~Seto,
  Phys.\ Rev.\ D {\bf 73}, 063001 (2006);
V.~Corbin and N.~J.~Cornish,
  Class.\ Quant.\ Grav.\  {\bf 23}, 2435 (2006).





\bibitem{Tinto:2001ii}
 M.~Tinto, J.~W.~Armstrong and F.~B.~Estabrook,
  Phys.\ Rev.\ D {\bf 63}, 021101 (2001).

\bibitem{prep}
  N.~Seto, in preperation



\bibitem{Flanagan:1993ix}
  E.~E.~Flanagan,
  Phys.\ Rev.\ D {\bf 48}, 2389 (1993).



\bibitem{Seto:2001qf}
  N.~Seto, S.~Kawamura and T.~Nakamura,
  Phys.\ Rev.\ Lett.\  {\bf 87}, 221103 (2001).


\bibitem{Kawamura}S. Kawamura, et al,\ 
 Class.  Quant. Grav. {\bf 23}, 125 (2006).

\bibitem{Kudoh:2005as}
  H.~Kudoh, A.~Taruya, T.~Hiramatsu and Y.~Himemoto,
  Phys.\ Rev.\ D {\bf 73}, 064006 (2006).




\end{thebibliography}
\end{document}